\documentstyle[prl,epsfig,twocolumn,aps]{revtex}
\begin{document}
\twocolumn[\hsize\textwidth\columnwidth\hsize\csname@twocolumnfalse%
\endcsname
\title{Kardar-Parisi-Zhang Equation And Its Critical Exponents Through  
Local Slope-Like Fluctuations. }  
\author{S.V. Ghaisas} 
\address{
 Department of Electronic Science, University of Pune, Pune 411007,
India}

\date{\today}
                  
\maketitle 

\begin{abstract} 
 Growth of interfaces during vapor deposition are analyzed on a 
discrete lattice. For a rough surface, relation between the roughness 
exponent $\alpha$, and corresponding step-step (slope-slope) couplings 
is obtained in (1+1) and (2+1) dimensions.  
From the discrete form and the symmetries of the growth problem, the 
step-step couplings can be determined. Thus $\alpha$ can be obtained. 
The method is applied to the linear EW as well as 
 to the non linear growth equation,  
the Kardar-Parisi-Zhang equation in all dimensions.  
Further, exact exponents for the fourth order linear and 
nonlinear terms relevant to the growth with non conserved noise are obtained
in (1+1) and (2+1) dimensions.
\end{abstract} 

] 
\narrowtext 
It has been shown recently \cite{sv1} that the kinetic considerations 
on a growing surface can lead to various growth terms in a growth 
equation representing a conserved growth from vapor. A 
growth equation helps in understanding the roughening of a growing interface.
It represents dynamics of height fluctuations due to the deposition noise.  
 The non-linear behavior in 
stochastic equations is of general interest. The mode coupling due to 
the non-linearity can lead to the nontrivial physical effects that are  
crucial in understanding many stochastic processes \cite{bar,kr2}.  
However, in the presence of non-linearity, absence of analytical solutions 
in many cases does not allow a conclusive understanding of the roughening 
phenomenon. In the present work,   
slope components are discretized ({\it i.e.} steps). 
The discrete form depends upon the height difference between the 
nearest neighbor sites. It represents step at a given site and 
is proportional to the local slope. In the 
presence of non conserved noise the height difference fluctuates
{\it within the correlation length}. We refer 
to these fluctuations as the step fluctuations. Coupling conditions 
between the step fluctuations can be determined from the 
discretization of the relevant terms in the growth equation 
and the symmetry associated with the 
 growth problem.   
 These conditions determine the exact asymptotic 
behavior of the growth. In particular, in a growth, roughness 
along the interface is measured in terms of the exponent $\alpha$ and the 
time evolution of the height-height (h-h) correlation in terms 
of the exponent $z$. Knowledge of $\alpha$ leads to the determination of 
$z$, in the case of conserved growth from the scaling behavior of noise 
and in the case of KPZ equation from the Galilean invariance \cite{kr2}.
The method is further applied to fourth order linear and non linear terms 
considered to be important in understanding the growth phenomenon. Since we 
obtain exact results the renormalization behaviour of non linear terms 
can be obtained without ambiguity.   

In the following we develop the method for obtaining the critical exponents 
for the second order growth equation, both conserved and non conserved. 
Same method is later extended for fourth order terms. 

 A linear equation representing 
interface motion  normal 
to the surface can be obtained in the frame of reference moving 
with the interface velocity by considering inter-planer hopping of adatoms
on the interface with a bias for downward or in-plane hopping toward step edge
 \cite{sv1}.   
It has the form 
\begin{equation}
    \frac{\partial h}{\partial t}=\nu_{0}\nabla^{2}h+\eta  
\end{equation} 
where, $\nu_{0}$ explicitly depends upon $F$, and $\eta$ is the noise 
due to the randomness in the deposition flux. It has the correlation given 
by $<\eta({\bf x},t)\eta({\bf x'},t')>=2D\delta({\bf x-x'})\delta(t-t')$. The 
angular brackets denote the average of the contents. Eq. (1) is known as 
Edward-Wilkinson (EW) equation \cite{ew}. The lowest ordered non-linear 
correction to EW equation was introduced by Kardar , Parisi , and Zhang 
\cite{kpz}. The resulting equation, 
\begin{equation}
    \frac{\partial h}{\partial t}=\nu_{0}\nabla^{2}h+ \lambda(\nabla h)^{2}
                     +  \eta  
\end{equation} 
 is known as KPZ equation. This is a non-conservative equation.  

  The KPZ equation has large number of applications in growth 
\cite{bar,kr2}. Its variants are also useful in analyzing various physical 
situations \cite{fish}. The directed polymer representation of the KPZ 
equation \cite{mh} is one of the most studied equation for understanding 
the non-linear dynamics of the interfaces. Being non-linear an exact solution 
of the KPZ equation is not possible analytically. However, many physically 
relevant quantities are obtainable from the analysis of the equation that 
can be measured experimentally. In particular the scaling exponents that 
characterize the growth can be measured experimentally. By applying scaling 
transformations, $x\rightarrow bx$, $t\rightarrow b^{z}t$, and $h\rightarrow 
b^{\alpha}h$ to the KPZ equation and demanding that the equation remains 
invariant under these transformations one obtains, $z=2$, $z+\alpha=2$, and 
$z=2\alpha+d$ for the equation in $d$ dimensions. Since KPZ equation is 
obtainable from the noisy Burger's equation \cite{bur,bar,kr2} using 
the transformation $v \rightarrow \nabla h$, where $v$ is the velocity field, 
Galilean invariance is implied in the KPZ equation. The coefficient $\lambda$ 
explicitly appears in the Galilean transformation rendering it constancy 
under renormalization \cite{kr2}. As a result , 

\begin{equation}
z+\alpha=2
\end{equation} 

 is the relation valid in all dimensions. One can measure $\alpha$ from the 
height-height (h-h) correlations, 
 
\begin{eqnarray}
G(x,t)&=&\frac{1}{N}\sum_{x'}(h(x+x',t)-h(x',t))^{2}\nonumber \\
&=&x^{2\alpha}f\left(\frac{x}{\xi(t)}\right)
\end{eqnarray}

where, correlation length $\xi(t)\sim t^{1/z}$. In the limit $x\rightarrow 
0$, $f\rightarrow 1$. Time exponent $\beta$, where $z=\alpha/\beta$ can be 
obtained by measuring the width over a substrate of length $L$ as, 
$w_{2}=\frac{1}{N}\sum_{x}(h(x,t)-\bar h)^{2} 
=L^{2\alpha}g\left(\frac{L}{\xi(t)}\right)$,
It can be shown that \cite{bar} for small times $w_{2}\sim t^{2\beta}$. 

In 1+1- dimensions $\alpha$ has been determined exactly using perturbation 
expansion \cite{kpz} and also using Fokker-Planck equation with fluctuation - 
dissipation theorem \cite{bar}. In 2+1- dimensions perturbation theory fails since 
renormalization process indicates existence of only strong coupling regime. 
The dimension $d_{c}=2$ is a critical dimension for KPZ equation. For $d>2$, 
renormalization reduces $\lambda$ to zero , signaling EW behavior in the weak 
coupling regime. The strong coupling regime has a different behavior than EW 
equation. Thus, for $d>2$, a phase transition is observed for KPZ equation
\cite{bar,kr2}. To obtain various exponents in dimensions $d\ge2$, 
non-perturbative methods are employed. There are computer simulations 
\cite{kk,mpp,ggr} and theoretical methods \cite{ml,other,cast}. 
Most of the theoretical 
approaches suggest that there is an upper critical dimension beyond which 
strong coupling vanishes indicating weak coupling EW like behavior. 
Simulations \cite{kk} suggest that there is no upper limit on the 
dimension. Using effective large distance field theory subject to a few 
phenomenological constraints, it is shown that $\alpha$=2/5 and 2/7 in 2+1  
and 3+1 dimensions respectively. This claim is not supported by large scale 
computer studies \cite{mpp}. It is shown \cite{mpp} that the values of 
$\alpha$ are not rational numbers as predicted in reference \cite{ml}. Using 
a pseudo spectral method \cite{ggr}, numerical solution of the KPZ equation 
in 2+1 dimensions is obtained. Based on this method, $\alpha$ is obtained 
from the saturated width ($0.37 \pm 0.02$), height-height correlation 
($0.38 \pm 0.02$) and structure factor ($0.40 \pm 0.02$).   
In the present work we obtain {\it exact} values of the exponents of the 
KPZ equation. This is done by discretization of the KPZ equation and from the 
symmetry requirements, evaluating the step-step couplings. 
A step fluctuation at site $i$ is defined to be $(h_{i}-h_{i+1})$. 
We will refer to it simply as 'step'. 
It is proportional to the local discrete slope at site $i$. 

 Consider a one dimensional substrate with a lattice constant $a$ such that 
$a>l_{d}$, where $l_{d}$ is the diffusion length. The physical lattice 
constant will be $\le a$. We consider growth on this substrate where atoms 
are depositing on the physical lattice and follow the relaxation rules as 
depicted by a conservative growth equation. In 1+1- dimensions such an equation can be written as, 

\begin{equation}
\frac{\partial h}{\partial t}=\frac{\partial}{\partial x}J+\eta
\end{equation}

where, $J$ is the local particle current at $x$. Eq.(5) is written in the 
frame of reference moving with the average velocity of the interface defined
by the incident flux $F$.  

 In 1+1- dimensions, the (h-h) correlations in the discrete form are 
\begin{equation}
G(n)=\left<(h_{i}-h_{i+n})^{2}\right>
\end{equation}
We assume that {\it the correlation length $\xi$ is very large compared to 
 the length $na$}. We define step at $i$ as
\begin{equation}
\delta x_{i}=h_{i}-h_{i+1}
\end{equation}
The local slope is then $-\delta x_{i}/a$.  
In terms of steps we have 
\begin{equation}
G(2)=<\delta x_{i}^{2}+\delta x_{i+1}^{2}
+2\delta x_{i}\delta x_{i+1}>. 
 \end{equation}
Let $<\delta x_{i}^{2}>=<\delta x_{i+1}^{2}>=\delta^{2}$ and , $<\delta x_{i}
\delta x_{i+1}>=s\delta^{2}$ where $s$ is the coupling between the 
steps around $i$. The distribution for $\delta x_{i}$ is {\it always}  
symmetric around zero and time independent \cite{suprgh}for an 
ensemble average.
 
 In the limit $\xi \rightarrow \infty$  Eq. (4) reduces to 
 $G(x)=cx^{2\alpha}$ where constant $c=G(1)$. Hence Eq. (8) can be written 
as 
\begin{equation}
2^{2\alpha}=2+2s
\end{equation} 
where $G(1)=\delta^{2}$ in the discretized case. Coupling $s$ uniquely 
determines $\alpha$. Thus, for $s$=-1/2,0,and 1, $\alpha$ is 0, 0.5, and 1 
respectively. 

We analyze the growth equations in 1+1- dimensions in the discrete form 
to obtain $\alpha$ for various terms. Consider EW equation, 
\begin{equation}
    \frac{\partial h}{\partial t}=\nu_{0}(\delta x_{i-1}-\delta x_{i})+\eta_{i}
\end{equation}
The first term on the right side represents second derivative in terms of 
local steps. In order that the steady growth continues , at any instant 
the surface must be characterized by the fluctuations in the second term
{\it i.e.} various values of second derivative.
These must be consistent with the requirement that $<\delta x_{i}^2>=
\delta^{2}$ and that corresponding distribution of 
$ \frac{\partial^{2} h}{\partial x^{2}}$ 
must be scalable. From the analysis of the linear second order growth 
equation \cite{bar,kr2} $\alpha=0.5$.   
 Hence $s$ must be zero for the EW 
equation. $s=0$ correspond to Gaussian distribution of slopes  
 with average fluctuation 
of $\delta^{2}$. The average squared difference between consecutive slopes 
is $\left <\left (\delta x_{i}-\delta x_{i-1}\right )^{2}\right >
=2(1-s)\delta^{2}$ and the average squared sum is
$\left <\left( \frac{\delta x_{i}+\delta x_{i-1}}{2}\right )^{2}\right >
=\frac{1}{2}(1+s)\delta^{2}$. For given $\delta^{2}$, the volume in the 
parameter space formed with these two parameters, squared difference and 
sum of consecutive slopes, decides the number of configurations available 
for the second derivative in Eq. (10). This volume is maximum for $s=0$.
This result can be generalized to any higher order term by noting that 
the steps in the present case are proportional to the particle current 
corresponding to the EW term. Thus by following above argument for any 
other conservative term in 1+1- dimensions , the condition for maximum 
number of configurations is same as $<J_{i}J_{i+1}>=0$.  
Thus, when all the values $-1/2\le s \le 1$ are scalable in 1+1- dimensions, 
in the presence of non conserved white noise the EW term drives the 
surface morphology to maximize the configurations for the linear second 
order term.     

For KPZ equation, the scaling transformations lead to $\alpha=1/3$, {\it 
i.e.} $s=-0.206299$. Renormalization is expected as a result of the 
non linearity. The KPZ exponents in 1+1- dimensions are obtained without 
resorting to perturbation approach \cite{bar}. Here we obtain by referring 
to the step-step coupling. For a scalable surface $\delta^{2}$ is constant. 
For a given $\delta^{2}$, each scalable value of $s$ will correspond to 
certain distributions of slopes on the surface. Any value of $s$ lying 
between -1/2 and 1 is possible under renormalization since Eq. (2) 
does not indicate any condition that will prefer any particular value of $s$. 
Under these conditions, the surface is expected to exhibit the {\it maximum 
range of fluctuations} in the local slope $\delta x_{i}$ consistent with 
the average of $\delta^{2}$. Fig. 1 displays the distributions of slopes for 
$s=0$ and $s=1$ with $\delta^{2}=1$. Distribution for $s=1$ is sharp and 
symmetric on both sides and for $s=0$ it is Gaussian. For non zero value 
of $s$, the average squared difference and sum of consecutive slopes 
constrain the fluctuations at a given site. {\it e.g.} for $s=1$ 
the average slope difference is zero while sum is $\delta^{2}$ giving 
the sharp distribution as in Fig. 1. For any scalable $s \ne 0$, a 
double peaked distribution is obtained. However, the Gaussian distribution 
provides the maximum range of slopes. Hence, for KPZ equation the 
renormalization drives the system toward $s=0$ with $\alpha=0.5$.   

 In 2+1- dimensions, we need to consider 
the steps in two directions. We define, $\delta x_{i,j}=h(i,j)-h(i+1,j)$, 
$\delta y_{i,j}= h(i,j)-h(i,j+1)$, 
$\delta x_{i-1,j}=h(i-1,j)-h(i,j)$, 
$\delta y_{i,j-1}= h(i,j-1)-h(i,j)$,....over a square lattice. 
We consider isotropic situation so that, 
$<\delta x_{i,j}^{2}>=<\delta y_{i,j}^{2}>=\delta^{2}$ and the linear 
coupling constant in the two directions is same, $s$.  
In 2+1- dimensions, apart from the linear couplings 
in the two directions, cross-coupling  $<\delta x_{i,j}\delta y_{i,j}>$ 
also contributes. $\alpha$ can be expressed in terms of the cross coupling 
through the following relation, 
\begin{eqnarray}
<(h(i,j-1)-h(i+1,j))^{2}>=<(h(i,j-1)-h(i,j))^{2}\nonumber \\
+(h(i,j)-h(i+1,j)^{2}\nonumber \\
+2(h(i,j-1)-h(i,j))(h(i,j)-h(i+1,j))\nonumber
\end{eqnarray}
  This can be written as, 
\begin{eqnarray}
G(\sqrt 2)=<\delta x_{i,j-1}^{2}>+<\delta y_{i,j}^{2}>\nonumber \\
+2<\delta x_{i,j-1}
\delta y_{i,j}>\nonumber 
\end{eqnarray}
 This leads to 
\begin{equation}
2^{\alpha}=2+2q
\end{equation}
where, $<\delta x_{i,j-1}\delta y_{i,j}>=q\delta^{2}$. 
For $-1/2 \le q \le 0$, $\alpha$ varies between 0 to 1. 
From Eqs. (9) and (11) we obtain 
\begin{equation}
s=1+4q+2q^{2}
\end{equation}

Consider a set of $N$ number of pairs of slopes such that the average cross 
coupling over this set is $q\delta^{2}$. One can characterize such a set 
by a distribution of slopes, along with 
a distribution of difference  and sum of the 
slopes. All these distributions are needed simultaneously to obtain the 
same coupling $q$ by averaging.    
 Same is true in the case of linear coupling.   
In the analysis to be followed we will consider {\it set of 
pairs of steps while averaging 
over cross or linear couplings}. 

To analyze various terms in 2+1- dimensions, we consider a two dimensional 
substrate with $N$ number of sites on a space renormalizable lattice. $N$ 
is large enough for averaging purpose. The growth is assumed to have reached 
steady state with correlations developed over large 
distances. Various configurations 
are developed on the surface in statistically significant number so that 
the spatial averaging can be considered to be 
equivalent to the ensemble averaging .  
In 2+1 dimensions given a reference height, morphology can be defined by 
specifying the $N$ pairs of slopes, one at every lattice point. In order to 
specify it in terms of linear couplings, a pair in x-
direction and a pair in y- direction are necessary . Thus a quadruplet of 
slopes at a given site is needed. The morphology defined by these 
quadruplets for a scalable surface is such that the average over linear 
step coupling is $s$ and that over the cross coupling is $q$ consistent 
with Eq. (12). However, the relevant growth term in the growth equation 
demands certain relationship between $s$ and $q$, which helps 
in obtaining the growth exponents.  
We analyze the EW equation in 2+1- dimensions. The EW term in discrete 
form is $(\delta x_{i-1,j}-\delta x_{i,j}+\delta y_{i,j-1}-\delta y_{i,j})$.  
Steps in this term form a quadruplet of 
steps at a given site $i$. Thus there 
are $N$ quadruplets defining the frozen EW surface. This surface represents 
solution of the Eq. (1).   
This set of steps can be considered 
to form $2N$ cross couplings $(\delta x_{i,j}\delta y_{i,j})$ and 
$(\delta x_{i-1,j}
\delta y_{i,j-1})$. The morphology of the surface is defined through 
the specification of set of quadruplet defined at every site, {\it i.e.} 
by specifying the value of $\nabla^{2}h$ at every point.   
Let the cross coupling constant be $q_{ew}$. This surface 
can also be described by linear couplings defined over the {\it same} set of 
quadruplets over $N$ sites. Let $s_{ew}$ be the linear coupling characterizing 
this set of $2N$ linear couplings $(\delta x_{i,j}\delta x_{i-1,j})$ 
and $(\delta y_{i,j}\delta y_{i,j-1})$ over $N$ number of sites.  
Let $2s_{ew}/2q_{ew}=1+p$ , where $-1<p$ so that 
$<\delta x_{i-1,j} \delta x_{i,j}>_{N}+<\delta y_{i-1,j} \delta y_{i,j}>_{N}
=-(1+p)(<\delta x_{i,j} \delta y_{i,j}>_{N}+<\delta x_{i-1,j} 
\delta y_{i,j-1}>_{N})$. 
 The subscript $N$ 
denotes the averaging is over $N$ sites and the negative sign ensures 
the directions in which the slopes are measured are consistent. 
This implies that,  the linear coupling $s_{ew}$ over the given 
surface representing solution of EW equation is same as cross 
coupling $q_{ew}$ on
another surface where slopes from the quadruplets defined over the 
original surface are scaled by $\sqrt{1+p}$ while keeping the 
$\delta^{2}$ constant. These two conditions cannot be simultaneously 
satisfied over the same original surface for any non zero $p$.  
 Hence $p$ must be zero.    
This implies $s_{ew}=q_{ew}$.  
 From Eqs. (11), (12) we get $\alpha=0$ for the EW term 
in 2+1- dimensions. Thus in 2+1- dimensional case, the surface morphology is 
characterized by scalable $\nabla^{2}h$ term for the condition $s=q$. 

Consider KPZ term in 2+1- dimensions. The discrete form is $(\delta^{2} x_{i,j})
+(\delta^{2}y_{i,j})$. This form suggests that {\it on the $N$ sites there are 
$N$ pairs of slopes} $(\delta x_{i,j},\delta y_{i,j})$. These pairs will be 
characterized by a cross coupling $q_{kpz}$. This set of pairs will have the 
appropriate rotational symmetry. The surface morphology 
is defined over $N$ sites with this set of $N$ pairs. Thus in 2+1- dimensions,
 over $N$ sites it is necessary and sufficient 
to have only $N$ pairs of slopes to define the morphology of the  
KPZ surface as compared to $2N$ in the case of EW surface.   
Corresponding to this surface 
there will be $2N$ linear couplings $(\delta x_{i-1,j} \delta x_{i,j})$ 
and $(\delta y_{i,j-1}  
\delta y_{i,j})$ each being defined over $N$ sites and characterized by 
$s_{kpz}$. We will consider set of $N$ quadruplets 
$(\delta x_{i-1,j}, \delta x_{i,j},\delta y_{i,j-1},\delta y_{i,j})$
 defined over the $N$ 
sites (same as in the case of EW term) 
over which these linear couplings are defined. The set of cross couplings 
is also defined over the same set of quadruplets since the symmetry of the 
problem allows any of the cross coupled pairs to be chosen 
from a given quadruplet. Let $2s_{kpz}/q_{kpz}=1+p$. As is discussed 
in the case of EW equation, we identify linear couplings $2s_{kpz}$ same 
as cross coupling $q$ on a surface where steps are scaled by a 
factor of $\sqrt{1+p}$ and $\delta^{2}$ remains unchanged.  
 This is possible only for $p=0$.Hence, $s_{kpz}=q_{kpz}/2$.  
 Using this relation,  
 Eqs. (11) and (12) we obtain   
\begin{equation}
\alpha=0.35702 \nonumber
\end{equation}
for this surface, accurate up to 5 digits. One can compute it with arbitrary 
accuracy. Note that for KPZ equation, the surface can also be defined using 
$2N$ pairs of slopes as in the case of EW surface. However, for KPZ surface 
this $2N$ pairs of slopes are sufficient to define 
a morphology but not necessary. 
This will lead to the well known result of EW surface for KPZ case in the 
weak coupling (in this case small $\lambda$ in Eq. (2)) limit. 

In 3+1- dimensions, we consider a cubic lattice. For the additional 
degree of freedom we define $\delta z_{i,j,k}=h(i,j,k)-h(i,j,k+1)$
as the step in the $z$ direction. The discrete form of the KPZ term is 
$(\delta^{2} x_{i,j,k})
+(\delta^{2}y_{i,j,k})+(\delta^{2}z_{i,j,k})$. There is a triplet of slopes at 
every site on the surface. It is required that average cross coupling 
over $N$ sites for any two directions must result in to same $q$ value, 
$q_{kpz}$. We can choose a set of $N$ pairs of cross coupling between
 $\delta x_{i,j,k}$ and $\delta y_{i,j,k}$ characterized 
by $q_{kpz}$. The slope 
$\delta z_{i,j,k}$ must be chosen so that its cross coupling 
with $\delta x_{i,j,k}$
 or $\delta y_{i,j,k}$ is $q_{kpz}\delta^{2}$. To satisfy this condition 
we consider at every site the average value $\frac{1}{2}(\delta x_{i,j,k}+
\delta y_{i,j,k})$ as one of the slopes in the cross coupling. The other slope 
is $\delta z_{i,j,k}$ chosen in such a way that corresponding coupling will 
result in to  a set of cross couplings that averages over $N$ sites to 
$q_{kpz}\delta^{2}$. This ensures the requirement that cross 
coupling between slopes in any two given directions is same, $q_{kpz}$.
Thus, $2N$ pairs characterized by $q_{kpz}$ are required to define the 
 morphology of a surface whose growth dynamics is governed by KPZ equation
 on a three dimensional surface.
Consider the linear couplings defined over this three dimensional surface. 
There are $3N$ linear couplings on the surface. As has been argued in the 
case of 2+1- dimensions, in the present case, on this KPZ surface only $2N$ 
cross couplings are characterized by $q_{kpz}$. Both the couplings are 
defined over the set of sextet of steps defined at $N$ sites. Following 
the arguments as in the case of (2+1) dimensional KPZ growth one arrives at   
the relation $s_{kpz}=\frac{2}{3}q_{kpz}$ on a three dimensional surface.  
 $s=(2/3)q$ gives $\alpha=0.28125$.

By continuing the argument for 2+1- and 3+1- dimensional growth it is 
possible to obtain the relation between $s_{kpz}$ and $q_{kpz}$ in any 
dimension. On a $d$ dimensional surface,  
$\alpha$ can be determined from the condition $s_{kpz}=\frac{d-1}{d}
q_{kpz}$ which provides next higher value of $\alpha$.  Thus
for KPZ equation,  
$\alpha$ can be determined in any dimension. This implies that there is 
no upper critical dimension for the KPZ equation. For EW term on a 
surface with dimension $d>1$, the condition is always $s=q$ giving 
$\alpha=0$. 

Next we consider the application to fourth order terms. It has been 
shown \cite{ldv} that the relevant non linearities to this order are 
$\nabla^{2}(\nabla h)^{2}$ and $\nabla \cdot (\nabla h)^{3}$. The 
corresponding linear term is $\nabla^{4}h$, referred as Mullin's term 
\cite{mu}. The application of perturbation up to second order to 
$\nabla^{2}(\nabla h)^{2}$ term shows that its coefficient is not 
renormalized \cite{dk1}. In the case of  $\nabla \cdot (\nabla h)^{3}$
it is shown to generate EW term \cite{dk2,kg}. 

In 1+1- dimensions $\alpha$ is unity for fourth order linear and 
$\nabla^{2}(\nabla h)^{2}$ term \cite{ldv,dk1}. In the later case it 
implies that the corresponding coefficient in the growth equation 
is not renormalized.  
It is 0.5 for  
$\nabla \cdot (\nabla h)^{3}$ term \cite{dk2}.  
Consider the quantity $<(\delta x_{i-1}-\delta x_{i})^{2}=2-2s$. 
It represents average {\it local} second derivative on the surface
and the difference in the local particle current for the EW term.
If this difference is not zero on the average, then corresponding 
term contributes in the growth evolution of the surface. In the present 
case it means that EW term is present if this difference is not zero. 
Hence for any other conservative 
term to be exclusively present in the growth equation, this current 
difference must be zero. It is zero for $s=1$ giving $\alpha=1$. 
Thus in 1+1 dimensions for $\nabla^{2}(\nabla h)^{2}$ and  $\nabla^{4}h$
terms $s=1$ \cite{bar,kr2}. The term $\nabla \cdot (\nabla h)^{3}$
is similar to EW term in that the current for this non linear term is 
related to EW current and preserves the sign of the current locally. 
Thus condition for maximum number of configurations for EW term is 
same as that for the non linear term. Thus for the  $\nabla \cdot 
(\nabla h)^{3}$ term $s=0$ or $\alpha=0.5$ \cite{dk2} in 1+1 dimensions. 
Thus in 1+1 dimensions with non conserved noise, $s=0$ and $s=1$ are the 
only two possibilities defining the surface morphology. 

In 2+1 dimensions as before we must consider linear and other than linear 
(otl) couplings. So far we have encountered cross coupling as one of the 
otl couplings. The main difference in the analysis of the higher order 
terms using these couplings compared to second order terms is that, 1) more 
than one lattice points are to be considered for defining the discrete 
form of the term and 2) otl couplings other than cross couplings are 
needed. The roughness exponent is obtained by equating the total minimum
linear couplings required to define the local morphology to the total 
minimum otl couplings along with the Eq. (12).

Consider the term  $\nabla^{2}(\nabla h)^{2}$ or ldv term,  
in 2+1- dimensions. In this 
case the corresponding discrete form is $\delta^{2}x_{i+1,j}+\delta^{2}
x_{i-1,j}-4\delta^{2}x_{i,j}+\delta^{2}x_{i,j+1}+\delta^{2}x_{i,j-1}
+\delta^{2}y_{i,j+1}+\delta^{2}y_{i,j-1}-4\delta^{2}y_{i,j}+\delta^{2}y_{i+1,j}
+\delta^{2}y_{i-1,j}$. There are ten steps required to define the 
local morphology of the surface. As can be seen from Fig.2, these steps 
are spread over total five lattice points. The morphology at every lattice 
point must be defined over five lattice points associated with the given 
lattice point. To define the local morphology in terms of linear couplings, 
we note that linear couplings at (i-1,j), (i,j+1), (i+1,j) and (i,j-1) are 
necessary. These contain total of 16 steps. However, the very nature of 
linear couplings is such that minimum two pairs of steps in orthogonal 
directions are needed at a given lattice point to allocate it the correct 
relative height. Thus the number of steps involved are in excess to the 
number required to define the local morphology. This situation is also 
encountered in the case of KPZ equation previously. The contribution of 
linear couplings is 8s. Let's consider otl couplings. Since the linear 
couplings are added, in the explicit form, all the steps, defining the 
morphology appear in the expression involving sum of eight different 
linear couplings in terms of the steps. Hence, while otl couplings are 
considered, the relative sign of all the couplings therein must be same. 
This ensures the presence of all the steps in the explicit expression 
for otl couplings. This means that otl couplings be formed between 
steps that are parallel or following. The example of a following type 
coupling is $<\delta y_{i-1,j}\delta x_{i,j}>$ where the coupling is 
formed between steps that are following (arrows) as in Fig. 2. This 
coupling is $q\delta^{2}$. To discriminate from the other cross coupling 
such as $<\delta y_{i,j}\delta x_{i,j}>$ which is $-q\delta^{2}$, we will 
call this coupling proportional to $-q$ as dispersing one. 
We have used this later type in 
arriving at the results for EW and KPZ equations. In the present case 
following otl couplings are considered. $<\delta y_{i-1,j}\delta y_{i,j}>$
, $<\delta y_{i+1,j}\delta y_{i,j}>$,  $<\delta x_{i,j-1}\delta x_{i,j}>$,
 $<\delta x_{i,j+1}\delta x_{i,j}>$,  $<\delta x_{i-1,j}\delta y_{i,j+1}>$,
 $<\delta y_{i,j-1}\delta x_{i+1,j}>$. There is 
another possibility of addition of five dispersing cross couplings from 
Fig. 2. These are $<\delta y_{i+1,j}\delta x_{i+1,j}>$, 
 $<\delta x_{i,j-1}\delta y_{i,j-1}>$,
 $<\delta x_{i,j+1}\delta y_{i,j+1}>$,  $<\delta x_{i-1,j}\delta y_{i-1,j}>$ 
and ,$<\delta x_{i,j}\delta y_{i,j}>$. These two possibilities are chosen 
because they are additive involving all the steps needed to define 
the  $\nabla^{2}(\nabla h)^{2}$ term. The couplings also are distributed 
to render correct weightages to the lattice points involved. Thus 
from Fig. 2 it is seen that points at (i+1,j) and (i,j+1) contribute 
three steps each while (i-1,j) and (i,j-1) contribute two each. 
Thus in both the sets, ratio of the couplings at (i+1,j) and (i,j+1) to 
those at (i-1,j) and (i,j-1) is 3:2. Any other additive possibilities 
do not render this ratio. It is further required that in the otl couplings 
height at (i,j) must appear. This requirement is that of locality of the 
term  $\nabla^{2}(\nabla h)^{2}$. Consider {\it e.g.}  $<\delta x_{i,j+1}
\delta y_{i,j+1}>$. This coupling does not contain $h(i,j)$. Such a 
coupling will independently couple with the steps at (i,j+1) leading 
to second order term. We will illustrate this point in connection with 
fourth order linear term.  
This requirement eliminates set of five dispersive 
couplings. As can be verified from the Fig. 2, all the couplings in 
other set involve $h(i,j)$.

In order to determine surface corresponding to the otl couplings 
preserving the locality, 
we need to find parallel coupling of the form  
$<\delta x_{i,j-1}\delta x_{i,j}>$. This is easily obtained by noting that 
$0=(h(i,j)-h(i+1,j)+h(i+1,j)-h(i+1,j-1)+h(i+1,j-1)-h(i,j-1)+h(i,j-1)-h(i,j))$.
Squaring this expression and writting in terms of steps gives the desired 
result, 
\begin{eqnarray}
 <\delta x_{i,j-1}\delta x_{i,j}>=(1+2q)\delta^{2} \nonumber
\end{eqnarray}
We denote this coupling as $q_{p}$.   
$<\delta x_{i-1,j}\delta y_{i,j+1}>$ can be obtained by noting that 
$5^{\alpha}=(\delta x_{i-1,j}+\delta y_{i,j}+\delta y_{i,j+1})^{2}/
\delta^{2}$. This leads to 
\begin{eqnarray}
<\delta x_{i-1,j}\delta y_{i,j+1}>=(
5^{\alpha}-3-2q-2s)/2) \nonumber
\end{eqnarray}
We denote this coupling as $q_{s}$. 
 We note that $\sqrt {5}=2^{1.160964}$.  
The relation between the linear and otl couplings 
is obtained from,  
\begin{equation}
8s=4q_{p}+2q_{s} 
\end{equation}
Using Eq. (11),(12) and (14) we obtain $q=-0.206$ giving $\alpha=2/3$ 
with possible numerical error beyond 4th digit. Thus for the ldv term 
present analysis shows that there are two possible surfaces, one with 
$\alpha=0.303$ and the other with $\alpha=2/3$. The scaling transformation 
gives $\alpha=2/3$ \cite{ldv}. This shows that the system has a fixed point 
where $\alpha=2/3$ which also happens to be the value obtained from the 
scaling transformations. Under these conditions, since noise is not 
renormalized under conserved growth conditions \cite{kr2}, the coefficient 
of the ldv term does not get renormalized. This shows clearly why 
perturbation analysis \cite{dk1} does not show renormalization even 
up to second order. Thus for ldv term the exact value of $\alpha$ and 
the one obtained from scaling transformations coincide. The situation 
in this respect is similar to any linear term.   

To obtain the roughness constant for  $\nabla^{4}h$
refer to Fig. 3. All the necessary steps contributing to the 
discrete form of this term are included in the diagram. By 
inspection of the Fig. 3, there are linear couplings at (i+1,j), (i-1,j), 
(i,j+1) and (i,j-1) as before contributing total of $8s$ to the 
linear coupling. It is interesting to verify the condition of locality 
and effect of its non observance in this case. From the Fig. 3, it is 
easily verified that one can obtain exactly eight cross couplings with 
six of them not involving the height $h(i,j)$. We immediately obtain the 
condition $s=q$ which is same as second order EW term. Thus it shows that 
the cross couplings independently get coupled to different lattice points
affecting heights at those points independently. This results in to 
effectively a single lattice point after discretization. However, single 
lattice point description corresponds to second order term. Hence, 
the coupling of $h(i,j)$ is necessary in otl couplings. Using the valid 
otl couplings, 
\begin{equation}
8s=8q_{p}+4q_{s}
\end{equation}
gives $q=0$ or $\alpha=1$ for the unique surface corresponding to the 
term $\nabla^{4}h$ \cite{bar}. 

Finally we consider the term  $\nabla \cdot (\nabla h)^{3}$. From the 
scaling transformations, in 2+1 dimensions, $\alpha=1/2$ for theis term.  
The steps 
included in its discrete form are shown in Fig. 4. The linear couplings 
contributing to the morphology are at (i,j) and (i+1,j+1) lattice points. 
Thus the contribution from linear couplings is $4s$. There are two 
possible sets of otl couplings in this case satisfying the condition of 
additive couplings, proper lattice point weightages and locality. 
First one will consist of 
$<\delta x_{i-1,j}\delta y_{i,j>}$, $<\delta x_{i,j}\delta y_{i,j-1}>$,
$<\delta x_{i,j}\delta x_{i,j+1}>$, and $<\delta y_{i,j}\delta y_{i+1,j}>$. 
This will contribute $4s=2q+2q_{p}$. This results in to $q=0.25$ or 
$\alpha=0.58497$. The other set is $<\delta x_{i-1,j}\delta y_{i,j}>$, 
$<\delta x_{i,j}\delta y_{i,j-1}>$,$<\delta x_{i,j}\delta y_{i+1,j}>$, 
$<\delta x_{i,j+1}\delta y_{i,j}>$. This will contribute $4s=4q$ 
thus leading to EW term \cite{dk1,kg}. Our analysis shows that this 
term has additional fixed point corresponding to $\alpha=0.58497$. 
The simulations and perturbation approach confirms to only EW behaviour
 \cite{dk1,dk2}. Significance of this additional fixed points needs 
further attention. 

We will now consider all the non linearities derivable from the non lnear 
terms considered above. Consider terms of the form $\nabla \cdot 
(\nabla h)^{2n+1}$ with n=1, 2, 3, ... . The relevant diagrams representing 
these terms in the discrete form are 
exactly same as that corresponding to the term with 
n=1. Hence, all such non linearities will have same behaviour {\it i.e.} 
like EW term \cite{kg} and with additional surface possible having 
$\alpha=0.58497$. Consider terms $\nabla^{2}(\nabla h)^{2n}$ with n= 1, 
2, 3, ... . In this case also, we find that the diagrams representing the 
discrete form are all same as that corresponding to n=1. Thus the 
behaviour of all such non linear terms is same as that corresponding 
to the ldv term. For non conservative terms such as $(\nabla h)^{2n}$ with 
n=1, 2, 3, ... we find that all such terms will behave as the KPZ term 
asymptotically. These results show that in the conserved or non conserved 
growth, the EW, KPZ and ldv are the only relevant terms up to fourth 
order. Further non linearities if relevant must contain higher number 
of derivatives than in ldv term such as $\nabla^{4}(\nabla h)^{2}$ etc.. 
Experimental values of exponents other than the terms 
discussed here are likely to be the result of 
correlated noise or transients in the growth. 
  
 In conclusion, we have shown that using the discrete form of the terms in 
a growth equation and the use of linear and cross coupling between steps 
at a site, $\alpha$ can be exactly determined for various terms in 1+1-
dimensions, and in 2+1- dimensions. In particular for KPZ term, 
the existence of $\alpha$ in all the higher dimensions shows 
that there is no upper critical dimension for the KPZ equation. The method 
can be  extended to any non linear stochastic equation with non conserved 
noise. Exact results for fourth order non linearities are obtained and 
show that the non renormalization for ldv term is a consequence of 
the coincidence of the exact value of $\alpha$ with that obtained from the 
scaling transformations. This study indicates that in the experiments 
the results are likely to be affected by the presence of the different 
non linear terms , the correlated noise and the transients.  

Acknowledgement: Author acknowledges useful suggestions by Prof. Joachim 
Krug of Koeln Univ., Germany.

\begin{figure} 
\epsfxsize=\hsize \epsfysize = 2.5 in
\centerline{\epsfbox{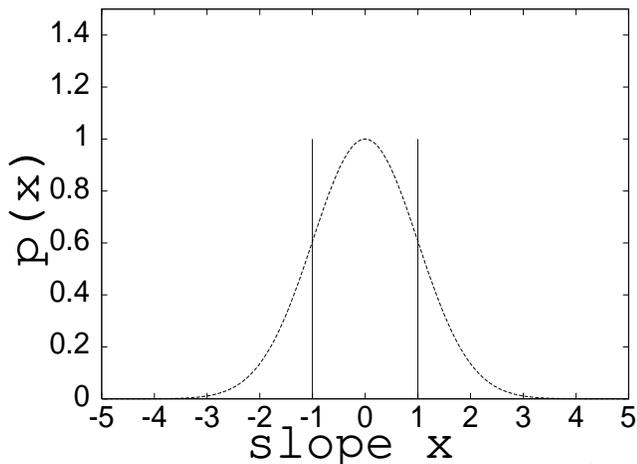}}
\caption{Slope distributions corresponding to $s=0$ (dotted curve) 
and $s=1$ (solid lines).   
}     
\label{fluc}  
\end{figure}
 
\begin{figure} 
\epsfxsize=\hsize \epsfysize = 2.5 in
\centerline{\epsfbox{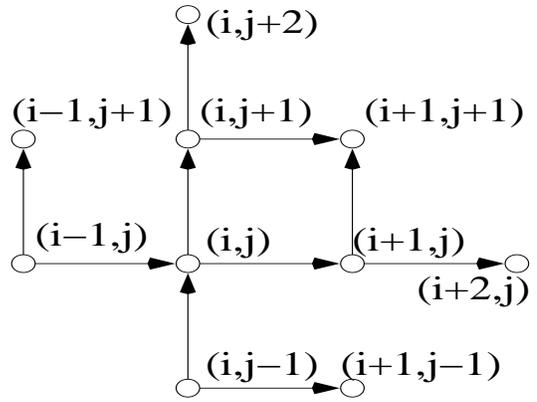}}
\caption{Steps contributing to the term $\nabla^{2}(\nabla h)^{2}$.
Each step is represented by an arrow as per the definition given in the text.   
}     
\label{fig2} 

\end{figure}

\begin{figure} 
\epsfxsize=\hsize \epsfysize = 2.5 in
\centerline{\epsfbox{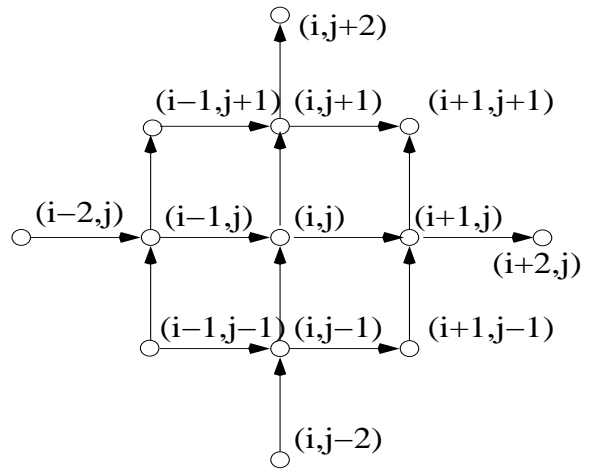}}
\caption{Steps contributing to the term $\nabla^{4}h$
}     
\label{fig3}  
\end{figure}

\begin{figure} 
\epsfxsize=\hsize \epsfysize = 2.5 in
\centerline{\epsfbox{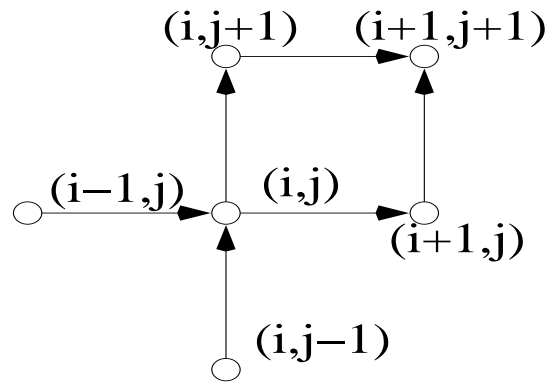}}
\caption{Steps contributing to the term $\nabla \cdot (\nabla h)^{3}$
}     
\label{fig4}

\end{figure}


\begin{references}
\bibitem{other} * Other 
references are available from the review articles like  
references 2) and 3) above.
\bibitem{suprgh}* In the cases where $\alpha>1$, $\delta^{2} $ can vary 
in time. This is referred as super rough growth and is discussed in reference 
3) above and references therein.
\end{references}
\end{document}